\documentclass[superscriptaddress,groupedaddress,nofootnoteinbib,11pt,notoc]{article}
\pdfoutput=1
\usepackage{jcappub}

\usepackage{amsmath, amsfonts, amsthm, amssymb, graphicx, color,hyperref,bbding}
\usepackage{subfigure}
\usepackage{dcolumn}
\usepackage{bm}
\usepackage{amssymb}
\usepackage{amsmath}
\usepackage{multirow}
\usepackage{float}
\usepackage{setspace}
\usepackage{booktabs}
\usepackage{enumerate}
\usepackage[table]{xcolor}
\definecolor{lightgray}{gray}{0.9}

\usepackage[utf8]{inputenc}

\def\ba{\begin{eqnarray}}
\def\ea{\end{eqnarray}}
\def\be{\begin{eqnarray}}
\def\ee{\end{eqnarray}}

\def\L{\mathcal{L}}

\def\stu{St\"uckelberg }
\def\Ost{Ostrogradsky }
\def\nn{\nonumber}

\def\d{\mathrm{d}}
\def\mn{_{\mu \nu}}

\def\({\left(}
\def\){\right)}

\def\mpl{M_{\rm Pl}}
\def\p{\partial}
\def\ie{{\em i.e. }}

\title{Ostrogradsky in Theories with Multiple Fields}
\author{Claudia de Rham and Andrew Matas}
\affiliation{CERCA, Department of Physics, Case Western Reserve University, 10900 Euclid Ave, Cleveland, OH 44106, USA}
\date{\today}

\abstract{
We review how the (absence of) Ostrogradsky instability manifests itself in theories with multiple fields. It has recently been appreciated that when multiple fields are present, the existence of higher derivatives may not automatically imply the existence of ghosts. We discuss the connection with gravitational theories like massive gravity and beyond Horndeski which manifest higher derivatives in some  formulations and yet are free of Ostrogradsky ghost. We also examine an interesting new class of Extended Scalar--Tensor Theories of gravity which has been recently proposed.
We show that for a subclass  of these theories, the tensor modes are either not dynamical or are infinitely strongly coupled.
Among the remaining theories for which the tensor modes are well--defined one counts one new model that is not field--redefinable to Horndeski via a conformal and disformal transformation but that does  require the vacuum to break Lorentz invariance. We discuss the implications for the effective field theory of dark energy and the stability of the theory.
}

\begin{document}
\maketitle
\flushbottom
\newpage

\section{Introduction}

Given the wealth of data on gravitational physics coming from cosmology with surveys such as Euclid \cite{Laureijs:2011gra} and the Large Synoptic Survey Telescope (LSST) \cite{Abate:2012za}, and the new era of gravitational wave astronomy that has been opened up by LIGO \cite{Abbott:2016blz}, it is important to develop theoretically consistent alternative theories of gravity to test against observations. This is especially timely as some modified theories of gravity can help to resolve important open problems in cosmology such as the nature of dark energy or can have distinctive signatures in inflation or in astrophysical observations \cite{Clifton:2011jh,Joyce:2014kja,Koyama:2015vza,Berti:2015itd}. From a theoretical perspective, modifications of gravity are interesting to study in their own right as they lead to insights on what properties a consistent theory must have.
\\

A major challenge in constructing consistent theories of gravity is avoiding the presence of an \Ost instability (or \Ost ghost) \cite{Ostrogradski} (see \cite{Woodard:2015zca} for a review of problems with the \Ost ghost and \cite{Chen:2012au} for methods to constructing theories without ghosts). The \Ost instability is a kinetic instability with an arbitrarily fast time scale, which can only be avoided if one includes new operators at the same scale as that of the ghost which remove the instability (\ie the interaction scale of a ghost should always be at least of the order of the cutoff of the theory). An \Ost instability may arise in generic modifications of gravity as they typically introduce higher derivative interactions which tend to excite the \Ost mode, unless great care is taken in choosing the form of these interactions. \\

Typically the \Ost instability is associated with the equations of motion involving third or higher order time derivatives. In the case of a single field, there is indeed a direct link between higher order equations of motion and the existence of an unstable, propagating \Ost mode. However when multiple fields are present, diagnosing an \Ost instability may be more subtle. \\

In the context of massive gravity this subtlety was for instance realized in the \stu language \cite{deRham:2011rn} and the helicity language in \cite{deRham:2011qq}, where the existence of higher derivatives is manifest beyond the decoupling limit. Historically, a problem with massive gravity has been the existence of the Boulware-Deser (BD) ghost mode \cite{Boulware:1973my}. In the decoupling limit, the BD ghost can be identified with higher order equations of motion for the helicity--0 mode \cite{Deffayet:2005ys,Creminelli:2005qk}. However, beyond the decoupling limit, the equations of motion for the \stu fields are higher order, even when the potential is chosen to avoid the BD ghost. To resolve this apparent paradox it is necessary to realize that the number of degrees of freedom is determined by the number of independent pieces of initial data that are needed to evolve the system. This cannot necessarily be read directly from the order of the equations of motion when multiple fields are present, because the equation of motion for one field can involve the derivative of the equation of motion of another field. This possibility is related to the existence of well--defined, invertible field redefinitions which can change the order of the equations of motion (but of course without changing the number of pieces of initial data that need to be specified).
\\

Precisely this same subtlety can be used in constructing consistent scalar--tensor theories. The most general class of scalar--tensor theories in which both the metric and the scalar field have second order equations of motion is known as Horndeski theory \cite{Horndeski:1974wa}, see also \cite{Deffayet:2013lga}. As scalar tensor theories, Horndeski theories have many important applications in cosmology ranging from inflation (see for example \cite{DeFelice:2011uc,Gao:2011qe,Kobayashi:2011nu}) to late universe cosmology (see for instance \cite{Deffayet:2013lga} and references therein).  \\

However just as is the case for massive gravity, demanding that the equation of motion for every field be second order may be overly restrictive. Indeed an explicit construction of such interactions has been performed and is known as `Beyond Horndeski' \cite{Gleyzes:2014dya,Zumalacarregui:2013pma}. Even though they were discovered later, Beyond Horndeski theories belong in the same class as the Horndeski theories in the sense that they are scalar--tensor theories avoiding an \Ost instability, and can be applied to phenomenology. It is then natural to wonder whether Beyond Horndeski is equivalent to Horndeski after a field redefinition. This interesting question was investigated in \cite{Gleyzes:2014dya,Zumalacarregui:2013pma,Gao:2014soa,Gleyzes:2014qga,Crisostomi:2016tcp}. In  \cite{Deffayet:2015qwa} it was shown that the higher derivatives that appear in the equations of motion of some of these theories can be eliminated and a Hamiltonian analysis for a specific model  showed the existence of a primary constraint. Then a fully systematic analysis for all Horndeski and Beyond Horndeski was performed in \cite{Langlois:2015skt} proving that the number of propagating degrees of freedom in these theories in three.  \\

Inspired by the existence of Beyond Horndeski, a very natural question in the context of scalar--tensor theories is: What is the broadest possible generalization of Horndeski which is free from the \Ost instability? Interesting progress along these lines was recently made by \cite{Langlois:2015cwa,Crisostomi:2016czh,Achour:2016rkg,Klein:2016aiq,Motohashi:2016ftl} leading to new classes of scalar--tensor theories endowed with an additional constraint that eliminate the ghost first found in \cite{Langlois:2015cwa}. Those theories have been dubbed {\it Extended--Scalar--Tensor theories of gravity}  (ESTs) \cite{Crisostomi:2016czh} or DHOST (Degenerate Higher Order Scalar-Tensor Theories) \cite{Achour:2016rkg}.\\

In this paper we review  some of the considerations that arise in analyzing the \Ost instability in theories with multiple fields.
First we present a general discussion about avoiding \Ost instabilities that emphasizes similarities between developments in massive gravity and in scalar--tensor theories. The fact that these similarities exist is perhaps not surprising as massive gravity can be viewed as a theory of gravity coupled to scalar fields forming a non--linear sigma model as discussed by \cite{deRham:2015ijs,deRham:2016plk}.\\

Then, we further analyze the nature of the degrees of freedom in these ESTs. While these theories have a primary constraint by construction and so the total number of propagating degrees of freedom is less than four (which would correspond to gravity in addition to a scalar field and its \Ost ghost), the next logical step is to determine exactly how these degrees of freedom are distributed among the scalar and tensor sectors. We find that in some cases, the tensor is either not dynamical (as also found in \cite{Langlois:2015cwa,Achour:2016rkg}) or infinitively strong coupled.\\

Among the remaining possibilities, one either counts Horndeski and field redefinitions from Horndeski as well as a specific EST model for which the tensor modes are well--behaved and show how the time derivatives of the lapse can always be removed in unitary gauge.  We discuss the implications for the construction of an effective field theory for dark energy \cite{Gleyzes:2013ooa,Gleyzes:2014rba,Gao:2014soa,Gleyzes:2015rua,Gleyzes:2015pma} (or for cosmology in general). \\

The rest of this work is organized as follows. In section \ref{sec:higher-derivative-example} we review techniques for diagnosing an \Ost ghost in theories with multiple fields and discuss their relevance for gravitational theories. In section \ref{sec:est-review} we then review the recently proposed EST theories and the existence of a primary constraint as well as the propagation of tensor modes. We then look at a special case of EST in section~\ref{sec:SpecialExample} which differs from Horndeski and beyond Horndeski.
We show how the presence of time derivatives on the lapse in unitary gauge can always be absorbed and discuss the implications for the construction of an effective field theory for cosmology. We also analyse the stability of this class of models and show the existence of gradient instabilities where no other interactions are present.  Finally we summarize our results  in section~\ref{sec:outlook}.

\section{Higher derivatives in Theories with Multiple Fields}
\label{sec:higher-derivative-example}

\subsection{Counting the number of Degrees of Freedom}
\label{sec:counting}

For a single field theory, the existence of higher--time derivatives leads to  the well--known \Ost instability \cite{Ostrogradsky} which manifests itself as an additional ghost degree of freedom (dof) which suffers from a kinetic instability and leads to an inconsistent theory (see Ref.~\cite{Woodard:2015zca} for the different consequences of this instability).\\

On the other hand, in theories with multiple fields, the existence of high--derivatives may not necessarily immediately lead to such an \Ost pathology. To our knowledge, one of the first explicit realization of this case manifested itself within the context of ghost--free massive gravity \cite{deRham:2010iK,deRham:2010kj} which was proven to be entirely free of ghosts in \cite{Hassan:2011hr,Hassan:2011ea} and yet higher--derivatives are manifestly present when considering interactions beyond the decoupling limit. The reason why higher--derivative are not necessarily fatal  in theories involving multiple fields (independently of their exact nature) was highlighted in \cite{deRham:2011rn,deRham:2011qq} and lies in the possibility to perform field redefinitions (rotating the field space variables) in a way that does not change the number of dof. Take for instance the two--scalar field toy--model in flat spacetime proposed in \cite{deRham:2011qq}
\ba
\label{eq:Toy1}
\L_{h \phi}=\frac 12 h \Box h + P(\phi, X) +\frac{\Box h (\p_\mu \p_\nu \phi)^2}{\Lambda^5}+ \frac{(\p_\mu \p_\nu \phi)^2 \Box (\p_\alpha \p_\beta \phi)^2}{2 \Lambda^{10}}\,,
\ea
(where later on $h$ may symbolically play the role of the gravitational field and $\phi$ the scalar field in a tensor--scalar theory of gravity) and $X=(\p \phi)^2$.  As written, the theory \eqref{eq:Toy1} manifestly involve higher derivatives. Yet, as highlighted in  \cite{deRham:2011qq}, \eqref{eq:Toy1} satisfies unitarity and all its scattering amplitudes are trivial.
Explicitly computing the scattering amplitudes for arbitrary N point functions and checking whether they satisfy unitarity is perhaps one of the most unambiguous way to determine whether the theory exhibits a ghost, but there exists a multitude of other ways to establish whether or not the theory \eqref{eq:Toy1} is free of any \Ost instability. The most  standard  way determine the number of dofs and the existence of ghosts is to perform a full Hamiltonian constraint analysis. In what follows we also present a few alternative tricks to establish the number of dofs.

\paragraph{1.-- Equations of motions and number of initial conditions:} One way to establish the absence of  \Ost instability for instance in \eqref{eq:Toy1} is to determine the number of initial conditions needed to solve the equations of motion.
 While higher derivatives are present in the equations of motion with respect to $\phi$,
\ba
\mathcal{E}_h &=& \frac{\delta \L}{\delta h}=\Box h + \frac{1}{\Lambda^5}\Box (\p_\mu \p_\nu \phi)^2\\
\mathcal{E}_\phi &=& \frac{\delta \L}{\delta \phi}=P_{,\phi}-2\p_\mu \(\p^\mu \phi P_{,X}\)+ \frac{2}{\Lambda^5}\p_\mu \p_\nu\left[\( \Box h  + \Box (\p_\alpha \p_\beta \phi)^2\) \p^\mu \p^\nu \phi \right]\\
&& \phantom{\frac{\delta \L}{\delta \phi}} = P_{,\phi}-2\p_\mu \(\p^\mu \phi P_{,X}\) +  \frac{2}{\Lambda^5}\p_\mu \p_\nu\left[\mathcal{E}_h\,  \p^\mu \p^\nu \phi \right],
\ea
no higher derivatives are present in the equations of motion with respect to $h$,
and all the higher derivatives in the equation of motion for $\phi$ actually disappear once solving the equation of motion for $h$.
This means that one only needs to specify two initial conditions per field to solve for the system, and the theory is not genuinely higher derivative. This counting is similar to that emphasized in \cite{deRham:2011rn,deRham:2014naa,Matas:2015qxa}. Notice however that in a theory with gauge symmetries one should also ensure that what should be auxiliary variables do not become dynamical in that process.

\paragraph{2.-- Field Redefinition:} Another way to see the absence of pathology is simply to perform the {\bf well--defined} and {\bf fully invertible} field redefinition
\ba
\label{eq:FieldDef}
h = \bar h -\frac{1}{\Lambda^5}(\p_\mu \p_\nu \bar \phi)^2\,, \qquad \phi = \bar \phi\,.
\ea
In terms of these new variables, the Lagrangian is manifestly healthy,
\ba
\L=\frac 12 \bar h \Box \bar h +P(\bar \phi,\bar X)\,.
\ea
We emphasize that  the existence of multiple fields is crucial in avoiding ghosts associated with higher derivatives. Indeed the field redefinition \eqref{eq:FieldDef} is only well--defined  because it corresponds to shifting the field $h$ by another field. On the other hand, another field redefinition of the form $\phi = \bar \phi + \frac{1}{\Lambda^5}(\p_\mu \p_\nu \bar \phi)^2 $ would not be fully--invertible and would hide non--perturbative dofs.

The existence of such a  field redefinition is simple to establish in this scalar--field toy--model but in a gravitational theory as in massive gravity or beyond Horndeski, it is highly non--trivial to perform these redefinitions and more systematic arguments can then be employed, including a full Hamiltonian analysis.

\paragraph{3.--  Hessian:} Another way to establish whether or not the theory \eqref{eq:Toy1} admits an \Ost ghost is to look at the Hessian of the dynamical variables as was argued in \cite{deRham:2011rn}. For simplicity and without loss of generality, we consider the ultra--local limit of the theory where both fields only depend on time. Since the field  $\phi$ enters with up to three time--derivatives we define two new variables $v$ and $w$  which are set respectively to $\dot \phi$ and  $\dot v$ with two Lagrange multipliers. The resulting Lagrangian then reads (after appropriate integrations by parts)
\ba
\label{eq:Toy12}
\L=P(\phi,\dot \phi^2) +\frac 12 \(\dot h+\frac{2}{\Lambda^5}w \dot w\)^2
+  \lambda_1 \(v-\dot \phi\)+\lambda_2\(w-\dot v\)  \,. \
\ea
It is now obvious that $h$ and $w$ do not have an independent conjugate momentum and this statement can be written more clearly by establishing the rank of the Hessian $\mathcal{H}_{ab}$ determined by
\ba
\mathcal{H}_{ab}=\frac{\p^2 \L}{\p \dot \Psi^a \p \dot \Psi^b}\,,
\ea
where the Lagrangian is written in first order form and the $\Psi^a$ represent all the fields involved, $\Psi^a=\{h, \phi, v, w\}$. One can easily check that in this case the rank of the Hessian is two, corresponding to two dynamical dof. We stress that the vanishing of the determinant of the Hessian only indicates the existence of a primary constraint. To remove a full dof, a secondary constraint should also be present. However without parity violation there cannot be half--integer number of dof, and therefore the existence of a secondary constraint is guaranteed in any theory which for instance preserves Lorentz invariance. In some of the theories which we will be looking at below, Lorentz invariance is broken and in these cases the existence of a secondary constraint is no longer necessary guaranteed.

\paragraph{4.-- Hamiltonian analysis:} Another direct and unambiguous way to establish whether or not the theory \eqref{eq:Toy1} admits an \Ost ghost is to perform a proper Hamiltonian analysis. However we emphasize that a Hamiltonian analysis is not the only way as we have shown through the previous arguments.

Once again, for simplicity and without loss of generality, we consider the ultra--local limit of the theory where both fields only depend on time. Since the field  $\phi$ enters with up to three time--derivatives we define two new variables $v$ and $w$  which are set respectively to $\dot \phi$ and  $\dot v$ with two Lagrange multipliers. The resulting Lagrangian then reads (after appropriate integrations by parts)
\ba
\label{eq:Toy12}
\L=P(\phi,v^2) +\frac 12 \(\dot h+\frac{2}{\Lambda^5}w \dot w\)^2
+  \lambda_1 \(v-\dot \phi\)+\lambda_2\(w-\dot v\)  \,. \
\ea
It is now obvious that $h$ and $w$ do not have an independent conjugate momentum.
In this language, $\lambda_{1,2}$ are auxiliary variables (Lagrange multipliers) while $h, \phi, v$ and $w$ are (in principle) dynamical variables with conjugate momenta
\ba
p_h=\frac{\p \L }{\p \dot h} = - \lambda_1 &\quad & p_{v_h} =  \frac{\p \L }{\p \dot v_h}\,.
\ea
As we have seen, the Hamiltonian is a very clean  way  to establish the number of dof but not the unique way. Moreover we emphasize that there are other relevant questions besides the number of dof, such as the scale at which degrees of freedom may enter, that may be easier to see in a different language.

\subsection{Setting  vs Decoupling}

As we have seen in the previous section, a theory which involves multiple fields can include higher derivatives without necessarily suffering from an \Ost instability. In what follows we shall emphasize the distinction between setting a variable to a given value and taking an appropriate decoupling limit. This distinction is important in the case of gravity where $h$ may for instance symbolically play the role of the metric (or the metric fluctuation about flat spacetime).\\

First we point out that taking a healthy theory with auxiliary variables and making these auxiliary variables dynamical is not a consistent procedure and can very well change the number of dofs. This confusion is at the origin of the results of \cite{Li:2015izu,Li:2015iwc} where the terms proposed manifestly exhibit a ghost as shown explicitly in \cite{deRham:2013tfa,deRham:2015rxa,deRham:2015cha,Matas:2015qxa}.\\

Similarly, the opposite procedure of `ignoring' the kinetic term of a field and {\it setting} it as fixed or considering it as an auxiliary variable would not be a consistent procedure and typically changes the number of dofs. For instance starting from the healthy theory \eqref{eq:Toy1} and simply setting for instance $h=0$  would lead to the following sick theory
\ba
\L_{h \phi} \xrightarrow{h := 0} \L_{\phi}= P(\phi,X) + \frac{(\p_\mu \p_\nu \phi)^2 \Box (\p_\alpha \p_\beta \phi)^2}{2 \Lambda^{10}}\,,
\ea
which has a ghost at the scale $\Lambda$.\\

Rather if one wants to decouple the two fields $h$ and $\phi$ it is instead possible to take a scaling limit which preserves the kinetic terms of the dynamical degrees of freedom while sending their interactions to zero. In the scalar field example \eqref{eq:Toy1}, we see that the scalar fields are already canonically normalized and scaling the interactions between the two fields corresponds to sending $\Lambda\to \infty$ so that the cubic term in \eqref{eq:Toy1} scales out. However in the same process we see that the last term also scales away and the resulting decoupling limit where the scalar field sees no interactions with $h$ (or what plays the role of the gravitational field in this example, \ie the scalar field sees flat spacetime in that limit) would be the theory which is manifestly well--defined:
\ba
\L_{h\phi}\ \xrightarrow{\ \Lambda \to\,  \infty\ }\  \L_{h\phi}^{\rm decoupled}=\frac 12 h \Box h+ P(\phi,X)\,.
\ea
Notice that taking this decoupling limit does not `kill' the interactions for $\phi$ itself that are present in $P(\phi,X)$. \\

Taking the decoupling limit argument from the other side, it means that if a theory is healthy, its decoupling limit ought to be healthy (\ie exhibit the correct number of degrees of freedom), and since in the decoupling limit the different fields do not interact, the single field \Ost argument should be valid and the theory should exhibit no higher derivatives when the coupling between the different fields are scaled to vanish. In the context of scalar--tensor theories this implies that in the limit where gravity decouples the scalar theory should end up being a generalized Galileon \cite{Deffayet:2009mn}, although taking that limit may not necessarily be trivial.  \\

A potential loophole behind the previous decoupling limit argument is if the theory itself necessarily breaks Lorentz invariance (not necessarily in its formulation but in its allowed vacua). This is what happens in the new class of scalar--tensor theories presented in \cite{Langlois:2015cwa} and further in \cite{Crisostomi:2016czh,Achour:2016rkg} which we review in what follows. One could then argue that by taking this road one could in principle allow the theory to have a preferred frame and hence allow theories which manifestly evade the \Ost ghost in a specific frame.

\section{Extended Scalar--Tensor Theories of Gravity}
\label{sec:est-review}

We now turn to Scalar--Tensor theories of gravity which involves a metric $g\mn$ and a scalar field $\phi$. We consider the Extended Scalar--Tensor Theories of Gravity (EST) or Degenerate Higher Order Scalar--Tensor Theory (DHOST) recently uncovered in \cite{Langlois:2015cwa} and analysed also in \cite{Crisostomi:2016czh,Achour:2016rkg} following some  of the logic established in \cite{Langlois:2015cwa,Crisostomi:2016tcp}. We follow most closely the conventions of \cite{Crisostomi:2016czh} and when overlapping, our results agree with those of   \cite{Achour:2016rkg}.
Following \cite{Langlois:2015cwa,Crisostomi:2016czh,Achour:2016rkg} we will consider the most general covariant action which is at most quadratic in second order derivatives on $\phi$ (and with no derivatives higher than two). The action takes the form
\ba
\label{eq:EST}
S_{\rm EST} = \int \d^4 x \sqrt{-g}\left( G(\phi,X)R+ P(\phi,X) + Q(\phi,X)\square\phi +  \sum_{i=1}^{5} A_i(\phi,X) \mathcal{L}_i \right)
\ea
with
\ba
\label{eq:EST-lags}
\mathcal{L}_1 &=&  \phi_{\mu\nu} \phi^{\mu\nu} \\
\mathcal{L}_2 &=&  (\square \phi)^2 \\
\mathcal{L}_3 &=&   (\square \phi) \phi^\mu \phi_{\mu\nu} \phi^\nu \\
\mathcal{L}_4 &=& \phi^\mu \phi_{\mu\rho}\phi^{\rho \nu}\phi_{\nu} \\
\mathcal{L}_5 &=&   (\phi^\mu \phi_{\mu\nu} \phi^\nu)^2\,,
\ea
 with the shorthand notation $\phi_\mu \equiv \p_\mu \phi$ and $\phi_{\mu\nu} \equiv D_\mu D_\nu \phi$, as well as $X \equiv (\partial \phi)^2$. The functions $G$ and $A_i$ $(i=1,\cdots,5)$  satisfy some relations which depend on the class of EST. \\

 If $G\ne 0$ to start with, then one can always go to Einstein frame and set $G\equiv 1$ however the relevance of $G$ becomes important when coupling to matter. An important aspect of these types of theories is therefore their stability when matter coupling to the metric $g$ is included. \\

All of the Lagrangians $\L_i$ are quadratic in $\phi_{\mu\nu}$ and each one of them leads to higher derivatives in the equations of motion. However there are special combinations of the $\L_i$ for which the equations of motion are second order in derivatives and in flat spacetime this corresponds to the quartic Galileon \cite{Nicolis:2008in,Deffayet:2009mn}.
 In addition when interaction with gravity is included \ie the theory has multiple fields, higher derivatives in the equations of motion are not necessarily fatal as illustrated in the previous examples and explained in \cite{deRham:2011rn,deRham:2011qq}.\\

This possibility was successfully exploited in \cite{Gleyzes:2014dya} where a family of `Beyond Horndeski' Lagrangians which have no ghost (at least without considering couplings to matter) was established.
Very recently, this possibility was pushed even further in \cite{Langlois:2015cwa} where is was shown even besides `Beyond Horndeski', that there are other classes of EST for which  the six functions $A_i, G$ satisfy special relations which allow the theory to enjoy a primary constraint which potentially removes the \Ost instability. For these new EST's, the functions $P$ and $Q$ can be freely chosen without affecting these conditions.  \\

One important point to stress though, is that almost none of thew new consistent EST/DHOST's proposed in  \cite{Langlois:2015cwa} and \cite{Crisostomi:2016czh} admit a Lorentz--invariant vacuum. We are therefore dealing with a new class of scalar--tensor theories for which the vacuum necessarily breaks Lorentz invariance and for which the scalar field $\p_\mu \phi$ is necessarily either spacelike or timelike and can never flip between the two. Since in the context of \cite{Crisostomi:2016czh} and \cite{Gleyzes:2014rba,Gleyzes:2014dya,Lagos:2016wyv} these theories were originally developed with cosmological applications in mind, it does make sense to think of them in vacua where the scalar field is timelike.

\subsection{Classes of EST/DHOST's}
First let us give a broad overview of the set of EST theories.
We will reproduce the exact relations defining these classes in Appendix \ref{app:est-details}. There are several different classes of EST which were identified in \cite{Langlois:2015cwa}. We will follow the naming scheme of \cite{Crisostomi:2016czh}.

The functions $P$ and $Q$ can be specified arbitrarily without introducing an \Ost instability. The EST theories are therefore defined by the relationships between the functions $A_i$ and $G$. The classes are

\begin{itemize}
\item {\bf Minimal Cases}: First there are, where $G=0$ (dubbed `Class M'). As we will see in sections \ref{sec:covariantExp} and \ref{sec:no-tensors}, the tensor modes have vanishing gradients in this case and are thus ill-defined. Further if $A_1=0$ the tensors are not dynamical fields.
\item {\bf Non--Minimal Cases}: The rest of the models all involve a non--vanishing $G$ and are called `Class N'. Among those,
\begin{itemize}
\item Class N-I satisfies $A_1=A_2$. This includes Horndeski and Beyond Horndeski. As shown in \cite{Crisostomi:2016czh}, this entire class can be generated by a field redefinition from Horndeski (up to a few subtle cases where the field redefinition may not be well-defined).
\item Class N-II: defined by $A_1=A_2=G/X$. Just like in the minimal cases, this class is ill-defined as it has no propagating tensor modes, as we will discuss in section \ref{sec:no-tensors}.
\item Class N-III: Finally there are theories where $A_1\neq A_2$. There are two subclasses, N-III-i which is a rather special case for which $A_1\ne G/X$ that we discuss in section \ref{sec:NIIIiExample} and N-III-ii for which $A_1=G/X$ and which again has no propagating tensor modes as we discuss in section \ref{sec:no-tensors}.
    \end{itemize}
\end{itemize}

In \cite{Achour:2016rkg} it was shown that each subclass below transforms into itself under a field redefinition of the form
\be
g_{\mu\nu} \rightarrow A(\phi,X)g_{\mu\nu} + B(\phi,X)\partial_\mu \phi \partial_\nu \phi.
\ee
Each class (except M-III) has 3 free functions. Therefore, we expect to be able to remove 2 of these 3 functions with the above field redefinition (except potentially for certain special cases where the field redefinition might fail to be invertible).

\subsection{Non--dynamical tensors}
\label{sec:covariantExp}

To gain some insight on the behaviour of some of these theories, we consider the following example (which belongs to the class M-III  considered in \cite{Langlois:2015cwa,Crisostomi:2016czh,Achour:2016rkg}, see \eqref{eq:MIII} in  Appendix \ref{app:est-details})
\ba
\label{eq:L2again}
\L_2 =\sqrt{-g} A_2 \, (\Box \phi)^2\,.
\ea
For simplicity we may consider $A_2$ to be a constant although none of the arguments below are affected by this choice.
Varying with respect to the metric and the scalar field we obtain the following strongly modified Einstein and Klein--Gordon equations:
\ba
\label{eq:eq2phi}
\mathcal{E}^{(2)}_\phi &=& \frac{1}{\sqrt{-g}}\frac{\delta \L_2}{\delta \phi}  =   \Box^2 \phi =0\\
 \mathcal{E}^{(2)}\mn &=& \frac{1}{\sqrt{-g}} \frac{\delta \L_2}{\delta g^{\mu\nu}}  =   \frac 12 g\mn \(\Box \phi \)^2 + \p_\alpha \phi \p^\alpha \Box \phi g\mn-2\p_{(\mu} \Box \phi \p_{\nu)}\phi=0\,,
 \label{eq:eq2uv}
\ea
where we symmetrize the indices as $V_{(\mu\nu)}=1/2 (V\mn+V_{\nu\mu})$.\\

From these equations of motion, it is now manifest that there are indeed fewer than four propagating dofs confirming the results of  \cite{Langlois:2015cwa,Crisostomi:2016czh}. However those dofs are not split into two tensor modes and one scalar mode. Rather the theory only contains a scalar mode. \\

To see this explicitly, we can consider the following equation:
\ba
\label{eq:eq2phi2}
\mathcal{E}^{(2)'}_\phi:\(g^{\mu\nu}-2\frac{\phi^\mu \phi^\nu}{X}\)\mathcal{E}^{(2)}\mn = 3(\Box \phi)^2=0\,.
\ea
As a result, in the vacuum, we need to have $\Box \phi\equiv 0$ and the rest of the dynamical equations are automatically satisfied.
We emphasize that this is an entirely covariant statement and is hence background independent, the only assumption has been the absence of matter fields as well as setting the functions $P(\phi,X)=Q(\phi,X)=0$.  If on the other hand we had chosen to include the stress--energy tensor for other matter fields on the right--hand side of \eqref{eq:eq2uv},  those would have appeared on the right hand side of \eqref{eq:eq2phi2} and the rest of the equations of motion could be read as constraints for the first derivative of the metric, but there would still  be no dynamical equations for the metric itself.

Having worked covariantly with a special example, we now turn to the more general classes of EST and perform an analysis of the tensor modes of FLRW to establish whether or not they are dynamical.

\subsection{Propagating tensors}
 \label{sec:no-tensors}

The EST theories were constructed to guarantee a primary constraint. Ideally this primary constraint will remove the Boulware-Deser ghost, leaving a healthy scalar sector, while leaving the dynamics of the tensor modes unchanged. However, a constraint analysis does not directly address this question. Therefore a natural first check is to understand how the tensor modes are affected by this constraint.\\

We now consider the general theory \eqref{eq:EST}.
So long as $G\ne 0$, this theory has an explicit Einstein--Hilbert term and we would expect tensor modes to be propagating, however as we shall see that this not always necessarily the case. Consider for instance the tensor perturbations $h_{ij}$ on FLRW so that the metric is $g\mn=\bar g^{\rm FLRW}\mn+h\mn$.
On this background solution, the quadratic Lagrangian for the tensor fluctuations is of the form
\ba
\L_{\rm tensor}=\frac {a^3 N} 2 \(\frac{1}{N^2} \left[G-X A_1\right] \dot h_{ij}^2+G\ h_{ij}\frac{\nabla^2}{a^2} h^{ij} - m^2_{\rm eff}(t)h_{ij}^2\)\,,
\ea
where $\nabla^2$ is the three--dimensional spatial Laplacian, $a$ is the scale factor, $N$ the lapse and $m_{\rm eff}$ is an effective mass term that depends on the background profile (\ie the background scalar field as well as the scale factor and the lapse). The exact expression for the effective mass term is not relevant to the discussion.\\

This result is easy to establish in FLRW but is actually much more general and background independent. Indeed so long as the scalar field is timelike we are always (at least locally) allowed to work in a gauge where $\phi=t$ (unitary gauge), and one can easily check that the previous results hold: \ie the kinetic term of the tensors is proportional to $(G-XA_1)$ (where in unitary gauge $X=g^{00}$) and the gradient terms are always proportional to $G$.
We therefore note two important cases
\begin{itemize}
\item Case 1: $G=X A_1$. This applies to the classes M-III, N-II, and N-III-ii. In this case  the tensor modes lose their kinetic terms. As a result, the tensor modes are not dynamical. This is in complete agreement with the other examples explored earlier where the EST considered lost the dynamical  tensor modes and consistent with the results presented in \cite{Langlois:2015cwa}.
\item Case 2: $G=0, A_1 \ne 0$. This applies to the classes M-I and M-II and in  particular to the isolated quartic Beyond Horndeski model which is not field redefinable back to Horndeski \cite{Crisostomi:2016tcp}. In this case the tensor modes still have a kinetic term, but no gradient terms. This implies that the tensor modes are infinitely strongly coupled and hence are ill--defined.
\end{itemize}
Note that if $\phi$ was instead spacelike, for instance $\phi=\phi(x)$, then the opposite would occur (the coefficient of the gradient along $x$ and that of the kinetic term of the tensors would switch) and the two previous cases would remain pathological. \\

As a result, we can conclude that N-III-i is the only remaining new class of theories which is potentially well--defined and not field--redefinable to a Horndeski theory using a conformal and disformal transformation.  We emphasize that N-III-i may still be field--redefinable to a more standard theory where the equations of motion are manifestly second order via a more general set of field transformations but finding the precise form of this more generic field transformation is in general difficult without more insight on how the theory is behaving. In what follows we shall explore this new theory N-III-i further and derive some implications for the construction of effective field theory for dark energy (or for cosmology in general).\\

\begin{table}[ht]
	\heavyrulewidth=.08em
	\lightrulewidth=.05em
	\cmidrulewidth=.03em
	\belowrulesep=.65ex
	\belowbottomsep=0pt
	\aboverulesep=.4ex
	\abovetopsep=0pt
	\cmidrulesep=\doublerulesep
	\cmidrulekern=.5em
	\defaultaddspace=.5em
\begin{center}
\begin{tabular}{c c c c}
            \toprule
 \multirow{2}{*}{Class}   \hspace{10pt}  & \hspace{10pt}
 \multirow{2}{*}{Tensors}  \hspace{10pt}  & \hspace{10pt}
 Lorentz  \hspace{10pt} & \hspace{10pt}
 Horndeski  \\
 &  &  invariant &
 (or field redefinition)\\[3pt]
    \midrule
 M-I,II,III & \XSolidBrush & \CheckmarkBold &  \XSolidBrush \\
 \midrule
 N-II &  \XSolidBrush & \CheckmarkBold &  \XSolidBrush \\
  \midrule
 N-III-ii &  \XSolidBrush & \XSolidBrush &  \XSolidBrush \\
 \midrule \rowcolor{lightgray}
  &   \CheckmarkBold  &  \CheckmarkBold  & \CheckmarkBold  \\
 \multirow{-2}{*}{ N-I}  & \XSolidBrush & \XSolidBrush   & \XSolidBrush \\
   \midrule \rowcolor{lightgray}
   N-III-i &  \CheckmarkBold & \XSolidBrush & \XSolidBrush\\
\hline
\end{tabular}
\end{center}
\caption{Different classes of Extended--Scalar--Tensor theories proposed in \cite{Langlois:2015cwa} and further in \cite{Crisostomi:2016czh}. Only the two subclasses have well--behaved tensor modes.
The first line of N-I corresponds to the models that are field--redefinable to Horndeski while the second line corresponds to the models for which such a field--redefinition would be singular.
When isolated from Horndeski (\ie with $G=0$), Beyond Horndeski (BH) belongs to the first class M-I or the second line of N-I. When coupled to Horndeski, BH is field redefinable to Horndeski and belong to the first line of N-I. Besides Horndeski and its field redefinitions, all the models that have well--defined tensors necessarily break Lorentz invariance.}
\label{tab:multicol}
\end{table}

\section{Implications for the Effective Field Theory of Dark Energy}

\label{sec:SpecialExample}

In the case of the new class of theories N-III-i, the vacuum necessarily breaks Lorentz invariance and the theory cannot make sense if $\p_\mu \phi$ is null. So within the entire region where the theory is defined, $\p_\mu \phi$ should either be time--like or space--like and either unitary gauge (\ie $\phi=t$) or the gauge $\phi=x^1$ can always be chosen everywhere (\ie everywhere where the theory makes sense). Moreover since the vacua of these theories necessarily break Lorentz invariance it is natural that there will be a preferred frame where all the equations of motion will be manifestly second order.
If, as in the case of the EST theories, we were primarily interested in theories relevant for cosmology (and potentially for dark energy), then focusing on backgrounds for which the scalar field is timelike is a justified restriction.\\

\label{sec:NIIIiExample}

As a specific example of the new class EST N-III-i, we shall consider the following theory
\ba
\label{Eq:ex}
\L = \mpl^2 \sqrt{-g}\(R -\frac{1}{2X}\L_2+\frac{2}{X^2}\L_4\)\,,
\ea
since $G=1$ and $A_1=0$ the tensor modes are well-behaved on FLRW.  Note however that this theory is only well-defined if $X$ is either positive or negative. In what follows we shall focus on the case where $X<0$. Once again, this implies that this theory admits no Lorentz invariant vacua.\\

While the equation of motion for the scalar field appears to be forth order in derivatives, it is easy to see that there exists a specific combination of the Einstein's equations which is only second order in derivatives on the scalar field and at most first order in derivatives on the metric:
\ba
\(X g^{\mu\nu}-2 \phi^\mu \phi^\nu\)\frac{\delta \L}{\delta g^{\mu\nu}}
= \mpl^2 \sqrt{-g} \(2 \L_1-\L_2+\frac{4}{X}\L_3-\frac{4}{X}\L_4-\frac{4}{X^2}\L_5\)\equiv 0 \,,
\ea
which is consistent with the arguments presented in section~\ref{sec:counting}. Since we are dealing with a gauge theory, to fully prove the absence of ghost here, one should also check that the rest of the equations of motion can be solved for the metric without involving time derivatives on the lapse and the shift. However since the existence of a constraint has already been proven fully covariantly in \cite{Langlois:2015cwa} and \cite{Crisostomi:2016czh}, for the rest of this argument it is sufficient to show that the constraint is indeed removing the \Ost ghost we may work for that in Unitary gauge. We also point out that since this model admits no Lorentz--invariant vacua, the existence of a secondary constraint is not necessarily guaranteed in principle, however this subtly is not an issue in this case.

\subsection{Unitary Gauge}

\label{sec:unitary}
To get more insight from this interesting new class of theories we work in unitary gauge where we can set the scalar field to be $\phi=t$. We further perform a $(3+1)$ ADM split \cite{Arnowitt:1962hi} so the metric is written as
\ba
\d s^2=-N^2 \d t^2+\gamma_{ij}\(\d x^i+N^i \d t\)\(\d x^i+N^j \d t\)\,,
\ea
and the standard Einstein--Hilbert term is given by (after integration by parts)
\ba
\sqrt{-g}R=N \sqrt{\gamma} \(R_3+[K^2]-[K]^2\)\,,
\ea
where $\gamma=\det(\gamma_{ij})$, $R_3$ is the three-dimensional curvature built out of $\gamma_{ij}$ and only involve spatial derivatives, square brackets represent the three-dimensional trace with respect to $\gamma_{ij}$ and $K_{ij}$ is given by
\ba
K_{ij}=\frac{1}{2N}\(\dot \gamma_{ij}-\nabla_{(i}N_{j)}\)\,.
\ea
In terms of these ADM variables the Lagrangian density of special example of EST N-III-i given in \eqref{Eq:ex} in unitary gauge is
\ba
\L=\mpl^2 N \sqrt{\gamma} \(R_3+[K^2]- [K]^2+\frac 12 \([K]+g^{0\mu}\p_\mu N\)^2+\frac{2}{N^2}g^{\alpha \beta}\p_\alpha N \p_\beta N\)\,.
\ea
A remarkable feature of this theory is the emergence of $\dot N^2$ terms. Since this theory was shown to be free of the \Ost ghost, it has been proposed that such operators be allowed in the general effective field theory for the description of dark energy. We emphasize that this is not the correct logic to constructing appropriate effective field theories. The only reason why operators involving $\dot N$ are allowed in this description is because
all such operators are removable with a field redefinition. In other words these operators all disappear after an appropriate change of variable and we are thus left with a theory in a much   `more conventional' form where no operators of the form $\dot N$ enter the field theory description in unitary gauge.\\

Indeed, by performing the following change of variable:
\ba
\label{eq:Fieldred}
\gamma_{ij}=\frac{1}{N^2}\tilde \gamma_{ij}\qquad {\rm and }\qquad N^i=\tilde N^i\,,
\ea
the Lagrangian density for \eqref{Eq:ex} is simply
\ba
\label{eq:Lunitarychange}
\L=\mpl^2 N \sqrt{\tilde \gamma} \(\tilde R_3+[\tilde K^2]-\frac 12  [\tilde K]^2+2 \(\frac{\tilde N^i\p_i N}{N^2}\)^2\)\,,
\ea
and involves no time--derivative neither on the lapse nor on the shift. The lapse is hence manifestly an auxiliary variable that can be integrated out while the shifts are the Lagrange multipliers ensuring three--dimensional diffeomorphism invariance.

\subsection{Field Redefinitions and Coupling to matter}

The field redefinition \eqref{eq:Fieldred} can be written covariantly as
\ba
g\mn =  \frac{- X}{\mpl^4} \tilde g\mn + \frac{B}{\mpl^4} \p_\mu \phi \p_\nu \phi\,,
\ea
where  $B$ is an arbitrary function (indeed in unitary gauge the disformal transformation generated by $B$ corresponds to a redefinition of the lapse). Setting $B=0$ for simplicity we have
\ba
g\mn = \frac{\sqrt{-\tilde X}}{\mpl^2} \tilde g\mn \,,
\ea
with $\tilde X=\tilde g^{\mu\nu}\p_\mu \phi \p_\nu \phi$,
leading to
\ba
\label{eq:exnewFrame}
\L=\sqrt{\tilde g \tilde X} \(\tilde R-\frac 12 \frac{\tilde  \L_2}{\tilde X}-\frac{\tilde \L_3}{\tilde X^2}+2 \frac{\tilde \L_4}{\tilde X^2}-\frac 12 \frac{\tilde \L_5}{\tilde X^3}\)\,,
\ea
which is manifestly well--behaved in unitary gauge as one can see from \eqref{eq:Lunitarychange}. \\

This result is neither new nor surprising. Indeed it was already pointed out in \cite{Gleyzes:2014rba} and \cite{Domenech:2015tca}, that the existence of $\dot N$ in unitary gauge does not necessarily imply the presence of ghost since those can be removable via a conformal and disformal transformations of the metric. However the points we would like to stress are the following:
\begin{enumerate}
 \item Operators involving $\dot N$ are actually {\it only} acceptable if they can be removed with a field redefinition and are hence not genuine operators that should be included in the effective field theory. In other words the existence of $\dot N$ in unitary gauge should not distract from the fact that $N$ should still remain an auxiliary variable and therefore arbitrary operators involving $\dot N$ cannot be introduced in the effective field theory.
  \item The metric for which all $\dot N$ disappear in unitary gauge (\ie $\tilde g\mn=(-X)^{-1}g\mn$) is the most natural metric matter should couple to covariantly and generic covariant couplings to the other conformally/disformally related metrics (for instance directly to the metric $g\mn$) could generically lead to ghosts as will be shown below.
\end{enumerate}
Interestingly for the class of theories N-III-i, there are no theories for which the tensors maintain a standard kinetic term (\ie $G\sim \mpl^2, A_1=0$) which do not involve time--derivative of the lapse in unitary gauge. However rather than reading this statement as opening up for the possibility of new operators in the effective theory of dark energy, we rather see this an indication that the tensor mode always manifest a peculiar kinetic structure (with a potentially time--dependent effective Planck scale). \\

Without coupling to matter the two theories  \eqref{Eq:ex} and  \eqref{eq:exnewFrame} are of course equivalent. However when coupling to matter the distinction between the two frames takes more significance. This is not so dissimilar to the distinction between Einstein frame and Jordan frame in standard scalar--tensor theories. While physics does not depend on the frame, much more insight on the stability of the theory can be gained from working in Einstein frame. This is because in Einstein frame the standard energy conditions on the matter sector can be used directly to imply the stability of the theory while in Jordan from those can take a modified shape. \\

To see this distinction for the EST's it is now instructive to couple the theory \cite{Langlois:2015cwa,Crisostomi:2016czh} to external matter, \ie other fields $\psi$. For instance let us  couple the theory   \eqref{Eq:ex}  to another scalar field $\chi$  which happens to have Galileon interactions,
\ba
\L = \mpl^2 \sqrt{-g}\(R -\frac{1}{2X}\L_2+\frac{2}{X^2}\L_4\) + \sqrt{-g}\(-\frac 12 (\p \chi)^2 + \frac{1}{\Lambda^3} (\p \chi)^2 \Box \chi\)\,.
\ea
Assuming for instance that there are solutions in unitary gauge for which $\chi=\chi(t)$ then in unitary gauge the theory can be written as
\ba
\label{eq:Lunitarymatter}
\L=\mpl^2 N \sqrt{\gamma} \(R_3+[K^2]- [K]^2+\frac 12 \([K]+g^{0\mu}\p_\mu N\)^2+\frac{2}{N^2}g^{\alpha \beta}\p_\alpha N \p_\beta N\)\nn \\
+ \frac{\sqrt{\gamma}}{N} \dot \chi^2\(\frac 12+\frac{\dot \chi}{\Lambda^3}\([K]+\frac{g^{0\mu}\p_\mu N}{N}\)+\frac{\ddot \chi}{N^2 \Lambda^3}\)\,.
\ea
After integrations by parts the terms going as $\dot \chi^2 \ddot \chi/N^3$ cancels that going as $\dot \chi^3 \dot N / N^4$ and the second line involves terms that go as $[K]\dot \chi^3$ but no terms that would involve a time derivative on the lapse (which of course precisely why Galileons lead to no ghost).
So it is now clear that the presence of $\dot N$ terms on the first line can no longer be removed by any field redefinition. To convince ourselves we could determine the determinant of the Hessian of the scalar modes defined as in \cite{deRham:2011rn} (up to order 1 coefficients),
\ba
\mathcal{H}_{IJ}=\frac{\p^2 \L}{\p \dot \Psi^I \p \dot \Psi^J}\sim\left(
\begin{array}{c c|@{}cc}
 \hspace{10pt} \mathcal{H}_{ab}^{\rm EST} \hspace{10pt}& & &  \( \begin{array}{c}
         \frac{\dot \chi}{\Lambda^3} \\
         0 \\
  \end{array}\) \\[15pt]
  \hline
  & & & \\[-5pt]
\( \begin{array}{cc}
         \frac{\dot \chi}{\Lambda^3}\  0 &
  \end{array}\) & & & 1+ \frac{\dot \chi}{\Lambda^3}  \\
\end{array}\right)\,,
\ea
with $\Psi^I=\{\gamma, N, \chi\}$ and $\Psi^a=\{\gamma, N\}$, \ie $\mathcal{H}_{ab}^{\rm EST}$ is the Hessian not including the matter field $\chi$.

It is straightforward to check that $\det \mathcal{H}_{ab}^{\rm EST}=0$ which is related to the existence of a constraint and the reason why the time derivatives on the lapse can be removed via field redefinitions. However  as long at the Galileon interaction is present (\ie finite $\Lambda$), the determinant of this full Hessian is now non--zero:
\ba
\det \mathcal{H}_{IJ} \sim \det \(\mathcal{H}_{ab}^{\rm EST} +\(
\begin{array}{cc}
\(\dot \chi/ \Lambda^3\)^2  & 0 \\
0 & 0 \end{array}\)\)\ne 0\,.
\ea
 Since the determinant of $\mathcal{H}_{ab}^{\rm EST}$ vanishes, the total  determinant of $\mathcal{H}_{IJ}$ is necessarily non--zero
 meaning that there are three scalar dynamical degrees of freedom corresponding to the Galileon, the scalar field $\phi$ and its \Ost ghost.  This is a simple consequence to the fact that general matter should not couple covariantly to the metric $g\mn$ (for which time derivative of the lapse enter in unitary gauge) but rather to the metric $\tilde g\mn$ (for which no time--derivatives of the lapse enter unitary gauge).\\

In conclusion while out of a healthy theory it is always possible to generate an infinite number of new formulations via field redefinitions. These fields redefinition may be of the conformal and disformal form but could also take on a much more complicated form, involving for instance higher derivatives of one field as was presented in the example \eqref{eq:FieldDef}. For instance one could generate a new infinite class of higher derivative theories which would be free from \Ost instability by starting from Horndeski and performing a change of variable where the metric is sent to a new metric which may involve many derivatives of the scalar field, so long as it does not involve more than one derivative on the spatial metric and no time derivative on the lapse or shift. In the vacuum this infinite class of theories would be fully equivalent to Horndeski.  However we stress that those field redefinitions matter when one couples to other fields. When seeking for consistent scalar--tensor theories relevant for cosmology it is hence `advisable' to focus (when possible) on those theories which do not involve time derivatives on the lapse in unitary gauge so that any covariant coupling to matter fields will preserve the constraint and the number of dof.

\subsection{Stability on flat FLRW}
\label{sub:stability}

Before concluding, we quickly glance at the behaviour of this exciting new class of models (N-III-i) by studying the scalar fluctuations on flat FLRW. We start by looking at the theory \eqref{eq:exnewFrame} in the vacuum. One could in principle add an additional arbitrary function of $P(\phi, X)$ as well as a generalized cubic Galileon, however we start by assuming that the interactions in \eqref{eq:exnewFrame} are the dominant ones and then draw conclusions for the general theory.
Then the Lagrangian for the background FLRW is simply
\ba
\L=-3\frac{a \dot a^2}{N^2}|\dot \phi|
\ea
where $a$ is the scale factor, $N$ the lapse and dots represent the time--derivatives.  It is then clear that in the vacuum, despite the presence of the scalar field $\phi$, the scale factor has to be constant $\dot a=0$ and the metric is Minkowski, while the profile for the scalar field remains undetermined signaling the existence of an accidental symmetry on that background. Note however that this background does break Lorentz invariance since $\phi$ ought to depend on time (indeed for $\dot \phi=0$ the tensors would lose their kinetic terms and that background would hence be infinitely strongly coupled). \\

Now turning to scalar fluctuations, and working in the gauge where $N=1$ and where the metric fluctuations are of the form
\ba
\label{eq:gauge}
\d s^2 = - (1+ 2 \Psi(t, \vec x)) \d t^2+ a^2(t)(1+2 \Phi(t, \vec x)) \d \vec x\,  {}^2\,,
\ea
and the scalar field is $\phi=\phi_0(t)+\delta \varphi(t,\vec x)$. The accidental symmetry present for the background manifests itself at the level of perturbations and the resulting perturbed Lagrangian is insensitive to $\Psi$:
\ba
\L^{(2)}=2a^3 |\dot \phi_0|\(- \dot \Phi^2+ \frac{1}{a^2}(\p_i \Phi)^2 \)+\frac{1}{2 a |\dot \phi_0|}\(\nabla^2 \delta \tilde \varphi\)^2\,,
\ea
where we have made the change of variables $\frac{1}{a^2}\nabla^2 \delta \tilde \varphi=\frac{1}{a^2}\nabla^2 \delta \varphi+|\dot \phi_0|\dot \Phi$. After integrating out $\delta \tilde \varphi$ we obtain the Lagrangian for single scalar degree of freedom as it should. As engineered in \cite{Langlois:2015cwa,Crisostomi:2016czh}, there is no \Ost instability for that field. However in that specific case we see that the field itself has the wrong sign kinetic term and is a ghost. This is no contradiction with  \cite{Langlois:2015cwa,Crisostomi:2016czh} and we confirm in this special example that the total of number of dofs is less than four as proved in \cite{Langlois:2015cwa,Crisostomi:2016czh} in all generality.\\

While the existence of a ghost instability on FLRW was shown  for a simple example, the existence of  instabilities -- and particularly gradient instabilities -- is actually generic to all class of N-III-i models proposed in \cite{Langlois:2015cwa,Crisostomi:2016czh} (which we recall is the only EST class of model for which the tensor modes are well--behaved and which is not field--redefinable to Horndeski). We show this generic statement in appendix~\ref{app:stability} where we look at all classes of   N-III-i models  {\it in the vacuum } and without any $P(\phi, X)$ or a generalized Galileon $Q(\phi, X) \Box \phi$.  If an instability occurs in the absence of matter and standard kinetic term or cubic Galileon, then the latter can `{\it save the day}' and restore stability if they dominate over the new interactions of the EST models. However if these EST interactions are never allowed to dominate, the fact that they are free of the \Ost instability is far less relevant since in the effective theory approach one can add any other interaction which may or may not carry an \Ost instability so long as these interactions do not dominate (see Refs.~\cite{deRham:2014fha,deRham:2014naa} for related discussions).

\section{Summary}
\label{sec:outlook}

While the existence of higher time derivatives automatically involve an \Ost instability, there has been a recent revived interest in how the \Ost instability manifest itself in theories with multiple fields where the notion of higher derivatives can be more subtle as it can change under field redefinitions (or mixing between different fields). In this paper we have reviewed different methods one can diagnose the existence of absence of \Ost instability in a theory with two scalar fields, and indicate that similar arguments can be applied to more involved theories such as those involving gravity and additional scalar fields, as in massive gravity or scalar--tensor theories of gravity. \\

With this general arguments in mind we have analysed an exciting new class of Degenerate--Higher--Order--Scalar--Tensor theories (DHOST) proposed in \cite{Langlois:2015cwa}, also known under Extended Scalar--Tensor--Theories (ESTs) investigated in \cite{Crisostomi:2016czh} (see also \cite{Achour:2016rkg}) and have discussed the implications for the effective field theory of dark energy. While it is true that in some frames, some of these theories involve time--derivatives on the lapse in unitary gauge, we emphasize that such operators are not free to enter independently in a consistent effective field theory. We also emphasize the importance of change of variables when it comes to coupling to matter and we study the stability of the new class of ESTs. We find that if we restrict ourselves to theories for which the tensors are well--behaved and the scalar is free from gradient or ghost instabilities on FLRW then the resulting ESTs reduce to Horndeski or field redefinitions thereof. However this is not to say that a subclass of the other theories could not provide an interesting phenomenology on more generic backgrounds, or when the standard kinetic terms, potential terms,  or the generalized cubic Galileon interactions dominate.

\vskip 10pt
\noindent {\bf Acknowledgments:}
We would like to thank Marco Crisostomi, David Langlois, Gianmassimo Tasinato, Andrew J. Tolley and Shuang-Yong Zhou for useful discussions. CdR and AAM are supported by a Department of Energy grant DE-SC0009946. CdR thanks the Galileo Galilei Institute for Theoretical Physics for its hospitality during the final stages of this work. 

\newpage

\appendix

\section{Explicit form of Extended Scalar Tensor Theories}
\label{app:est-details}
In this appendix we review the class of DHOSTs/ESTs introduced in \cite{Langlois:2015cwa} and follow the conventions and notations of \cite{Crisostomi:2016czh}. All the results presented here were derived in those papers and we simply present them here for completeness. Each class of EST has 3 different free functions (except for M-III which is special). In principle, we expect to be able to reduce two of those functions using a combination of conformal and disformal transformations.\\

First we recall that the general EST action is given by
\be
S = \int \d^4 x \sqrt{-g}\(G(\phi, X)\, R + \sum_{i=1}^5 A_i(\phi,X) \mathcal{L}_i \)
\ee
where the $\mathcal{L}_i$ are given in Equation \ref{eq:EST-lags}. We now look at special cases that amount to constraints we can impose on the $A_i$. In what follows, $G_X$ denotes the partial derivative of $G$ with respect to $X$.

\subsection{Minimal Cases ($G=0$)}
\noindent
\textbf{M-I}\\
The free functions are $A_1,A_2,A_3$ (with the restriction $A_2 \ne -A_1/3$). The other functions are given in terms of these by
\ba
A_4 = -\frac{2A_1}{X} \nn \quad {\rm and}\quad
A_5 = \frac{4 A_1 (A_1+2 A_2)-4 A_1 A_3 X+3 A_3^2 X^2}{4 (A_1 + 3 A_2) X^2}
\ea

\noindent
\textbf{M-II}\\
The free functions are $A_1, A_4,A_5$. The other functions are given by
\ba
A_2 &=& -\frac{A_1}{3} \nn \\
A_3 &=& \frac{2}{3} \frac{A_1}{X}
\ea

\noindent
\textbf{M-III}\\
This case is special and is defined by
\ba
\label{eq:MIII}
A_1=0
\ea
and $A_2,A_3,A_4,A_5$ are free. Note that since $G=0$ that $A_1=G/X$ for this case and therefore M-III is actually a special case of N-III-i.

\subsection{Non-minimal Cases ($G\ne 0$)}
\noindent
\textbf{N-I}\\
The free functions are $G,A_1, A_3$, subject to the condition that $A_1 \ne G/X$. The other functions are given by
\ba
A_2 &=& -A_1  \\
A_4 &=& \frac{1}{8(G-A_1 X)^2} \Big( 4 G\(3(A_1 - 2 G_X)^2 - 2 A_3 G \) - A_3 X^2(16 A_1 G_X + A_3 G) \nn \\
&&+4 X \(4 X\(3 A_1 A_3 G+16 A_1^2 G_X - 16 A_1 G_X^2 - 4A_1^3 + 2 A_3 G G_X\)\) \Big)  \\
A_5 &=& \frac{1}{\(G-A_1 X\)^2}\(2 A_1 - A_3 X - 4 G_X\)\( A_1 \(2A_1 + 3A_3 X - 4 G_X\) - 4 A_3 G \)
\ea

\noindent
\textbf{N-II}\\
The free functions are $G,A_4, A_5$. The other functions are given by
\ba
A_1 &=& \frac{G}{X} \nn \\
A_2 &=& -A_1 \nn \\
A_3 &=& \frac{2}{X^2}\(G-2X G_X\)
\ea

\noindent
\textbf{N-III-i}\\
The free functions are $G,A_1, A_2$, subject to the conditions that $A_1 \ne -A_2$ and $A_1 \ne G/X$. The other functions are given by
\ba
A_3 &=& 4\frac{G_X}{G}\(A_1 + 3 A_2\) - \frac{2}{X} \(A_1 + 4A_2 - 2 G_X\) - \frac{4 G}{X^2} \nn \\
A_4 &=& 2\frac{G}{X^2} + 8 \frac{G_X^2}{G}- \frac{2}{X}\(A_1 + 2G_X\)\nn \\
A_5 &=& \frac{2}{G^2 X^3}\Big( 4G^3 + G^2 X\(3 A_1 + 8A_2 - 12 G_X \) + 8 G G_X X^2\(G_X - A_1 - 3 A_2\) \nn \\
&& +6 G_X^2 X^3 \(A_1 + 3 A_2 \) \Big)
\ea

\noindent
\textbf{N-III-ii}\\
The free functions are $G,A_2, A_3$, subject to the condition that $A_2 \ne -G/X$. The other functions are
\ba
A_1 &=& \frac{G}{X} \nn \\
A_4 &=& \frac{8 G_X^2}{G} - \frac{4G_X}{X} \nn \\
A_5 &=& \frac{1}{4 G X^3\(G+A_2 X\)}\Big( 4G^3 + G^2 X\(3A_1 + 8A_2 - 12G_X\) \nn \\
&& + 8 G G_X X^2 \(G_X - A_1 - 3A_2\) + 6 G_X^2 X^3\(A_1 + 3 A_2\)  \Big)
\ea
Note that the constraint for $A_4$ is the same in both N-III-i and N-III-ii (taking into account that $A_1 = G/X$ in the latter case).

\section{Perturbations for N-III-i about flat FLRW}
\label{app:stability}

In this appendix we study the interesting class of N-III-i models. As shown in \cite{Crisostomi:2016czh}, as well as in section~\ref{sec:unitary}, those models may involve time--derivatives of the Lapse in unitary gauge but those can always be absorbed via field redefinitions. In order to focus the analysis, it is therefore sufficient to restrict ourselves to the subclass of models which involves no time--derivatives on the lapse in unitary gauge without needing to perform a field redefinition. \\

First we notice that any theory in the class N-III-i for which the function $G$ is of the form
\ba
G(\phi, X)=\sqrt{-X}\tilde G (\phi)\,,
\ea
(where $\tilde G$ could be an arbitrary function of the field),  involves no time--derivatives on the lapse in unitary gauge\footnote{To be able to go to unitary gauge in the first place, the scalar field should be time--like and we therefore choose the branch where $X<0$.} (without needing to resort to any field redefinition). In order to focus on the stability of the theory and avoid unnecessary change of variables, we therefore restrict ourselves to the sub--class of N-III-i EST theories for which we have
\ba
\label{eq:G}
G(\phi,X)=\sqrt{-X}\,.
\ea
Indeed even if we were dealing with a theory of the class N-III-i for which $G$ is different to start with, we can always put $G$ in the form of \eqref{eq:G} via an appropriate conformal transformation and as shown in \cite{Crisostomi:2016czh} such a transformation would map a N-III-i theory into another N-III-i theory, so setting $G$ as in \eqref{eq:G} amounts to no loss in generality (unless we started coupling to matter in which case those conformal transformations would matter, but as argued in section \ref{eq:Lunitarymatter} when one couples to matter it is usually `wiser' to do so to the frame which involves no time derivatives on the lapse in unitary gauge.)\\

Notice that the functions $\tilde G(\phi)$ is a priori arbitrary. However such a function can also always be set to unity by appropriate conformal transformation which only involve the scalar field and not its derivatives. Such a conformal transformation could generate a kinetic term for the scalar field but as mentioned in section~\ref{sub:stability} for the purpose of this discussion we focus on the case where only the EST interactions are present and we can therefore simply set $\tilde G(\phi)=1$. If $\tilde G'(\phi)\ne 0$ is needed to ensure stability, this signals the fact that the kinetic term for $\phi$ one would get after the conformal transformation needs to dominate over the EST interactions, and there is hence less motivations in ensuring that the EST interactions are ghost--free if they are never allowed to dominate.\\

Following the previous arguments and those of section~\ref{app:stability}, in the rest of this section  we consider the theory in the vacuum and in the absence of $P(\phi,X)$ terms and generalized cubic Galileon. We denote by $\mathcal A$ the following function:
\ba
\mathcal A(\phi, X)\equiv -2 (-X)^{3/2}\(A_1+3 A_2\)\,,
\ea
as this function enters frequently in the analysis. \\

Starting with $G=\sqrt{-X}$ and arbitrary functions $A_{1,2}$ we see that the resulting theory is very similar to what was derived in section \ref{sub:stability}. At the background level, the Lagrangian on flat FLRW is
\ba
\L=-6 \frac{a \dot a ^2 |\dot \phi_0|}{N^2}\(1-\frac {\mathcal A}2 \frac{|\dot \phi_0|}{N}\)\,.
\ea
The background equations of motion impose either:
 \begin{enumerate}[\quad (a)]
 \item $a=$ const,
 \item or in general admits another branch of solution with the following constraint:
\ba
\label{eq:CaseIcons1}
\mathcal{A}_{,X}=1\,.
\ea
In the special case of section \ref{sec:SpecialExample}, $\mathcal{A}$ was linear in $X$ so we could never be in that branch of solution since that equation could never be satisfied.  However for generic functions $\mathcal{A}$, the constraint \eqref{eq:CaseIcons1} could admit consistent solutions.
\item Finally, in principle we could also have the solution $\dot \phi_0=0$ but the tensors would be ill--defined on that background, so we do not consider it any further.
\end{enumerate}

In what follows, we analyse the two first cases one after the other.
\begin{enumerate}[(a)]
\item  Starting with the branch where $a=$ const in the background, we find that the resulting Lagrangian for the scalar field fluctuation (working in the same gauge as \eqref{eq:gauge}) is (after integrating out the constraints),
\ba
\label{eq:Phi}
\L_{\Phi}=2\frac{a^3}{\dot \phi_0^2}(G-X A_1)\(3\dot \phi_0^2- 2\frac{G-XA_1}{A_1+A_2} \)\dot \Phi^2-2 a G \Phi \nabla^2 \Phi\,,
\ea
and that branch of solutions always exhibits a gradient instability (if $G>0$, the instability is in the scalar mode and had we taken $G<0$, the instability would have been in the tensor modes). \\

\item Next we may look at the branch which satisfies \eqref{eq:CaseIcons1} for the background. The constraints are slightly more intricate in that case but an be solved for\footnote{The result in \eqref{eq:Psi} holds for $\mathcal{A}_{,XX}\ne 0$. If we assume instead that $\mathcal{A}$ is linear in $X$ then the constraint \eqref{eq:CaseIcons1} can only be satisfied if $\mathcal{A}=\mathcal{A}_0(\phi)+X$. In that case $\Psi$ entirely disappears from the Lagrangian  as was the case in section~\ref{sub:stability} and the rest of the stability analysis presented below remains entirely valid.}
    \ba
    \label{eq:Psi}
    \Psi=\frac{\delta \varphi}{2|\dot \phi_0|}\frac{\mathcal{A}_{,X\phi}}{\mathcal{A}_{,XX} }-\delta \dot \varphi\,,
    \ea
 After integrating out $\Psi$, it is easier to perform the change of variable,
\ba
\delta \varphi=\frac{a}{\dot a}|\dot \phi_0| \Phi+\chi,
\ea
(notice that if we had $\dot a=0$, we could go back to the first case scenario). Finally we can integrate $\Phi$ out. Its exact expression is not particularly illuminating, but the resulting Lagrangian density for the scalar degree of freedom $\chi$ is
\ba
\L_{\chi}=\frac{2 a \dot a^2}{\dot \phi_0^4}(G-X A_1)\(3 \dot \phi_0^2-2 \frac{G-X A_1}{A_1+A_2}\)\dot \chi^2-2 \frac{\dot a^2}{a |\dot \phi_0|}\chi\nabla^2 \chi\,,
\ea
which is remarkably similar to that found in the previous branch of solution \eqref{eq:Phi}. Just like in that case, as long as $3 \dot \phi_0^2-2 \frac{G-X A_1}{A_1+A_2}>0$, the time kinetic term would have the correct sign, however the gradients always enter with the wrong sign and just like for the previous branch, this class of theories suffer from gradient instabilities on flat FLRW.
\end{enumerate}

\newpage

\bibliographystyle{JHEPmodplain}
\bibliography{refs}

\end{document}